\documentclass{article}

\usepackage[preprint]{neurips_2025}
\usepackage{natbib}
\setcitestyle{authoryear,open={(},close={)}}
\usepackage{fancyhdr}
\usepackage[utf8]{inputenc} 
\usepackage[T1]{fontenc}    
\usepackage{hyperref}       
\usepackage{url}            
\usepackage{booktabs}       
\usepackage{amsfonts}       
\usepackage{nicefrac}       
\usepackage{microtype}      
\usepackage{xcolor}         
\usepackage{hanging}        
\usepackage{graphicx}       
\usepackage{placeins}
\usepackage{enumitem}
\setlist[itemize]{leftmargin=0.5em}

\title{Military AI Cyber Agents (MAICAs) Constitute a Global Threat to Critical Infrastructure}

\author{%
  Timothy R. Dubber\\
  MINT Lab\\
  Australian National University\\
  Canberra, ACT 2601 \\
  \texttt{timothy.dubber@anu.edu.au} \\
  \And
  Seth Lazar\thanks{Authors statement: Dubber led this project under Lazar's supervision. Ideas from the paper were originally presented at the MILA Harms and Risks of AI in the Military 2024 Conference, Montreal. We acknowledge funding from Australian Research Council grant FT210100724 and the Templeton World Charity Foundation.} \\
  MINT Lab\\
  Australian National University\\
  Canberra, ACT 2601 \\
  \texttt{seth.lazar@anu.edu.au} \\
}

\begin{document}

\maketitle

\begin{abstract}
This paper argues that autonomous AI cyber-weapons---Military-AI Cyber Agents (MAICAs)---create a credible pathway to catastrophic risk. It sets out the technical feasibility of MAICAs, explains why geopolitics and the nature of cyberspace make MAICAs a catastrophic risk, and proposes political, defensive-AI and analogue-resilience measures to blunt the threat.

\end{abstract}

\section{Introduction}
 \vspace{-0.5\baselineskip}

AI risk advocates and military AI ethicists alike have paid too little attention to the loss-of-control risks posed by Military AI Cyber Agents (MAICAs): fully autonomous AI agents that plan and execute cyber operations. Nation-states must forestall risks associated with MAICAs’ inevitable development by: 1. taking counter-proliferation steps to avoid MAICAs falling into the hands of rogue actors; 2. conducting defensive research to develop the counter-AI capabilities necessary to degrade a rogue-MAICA; and 3. investing in critical infrastructure resilience and analogue redundancy to limit the damage of a rogue MAICA.

\section{The risk of AI powered cyberwarfare is underappreciated }
 \vspace{-0.25\baselineskip}

In this paper, we argue that MAICAs represent a credible path to catastrophic risk, which we define as \textit{"events of low or unknown probability that if they occur inflict enormous losses often having a large non-monetary component."} \citep{Posner2008-vm}.\footnote{``The Indian Ocean tsunami of 2004 is at the lower level of the catastrophic-risk scale of destruction; examples from higher levels including large asteroid strikes, pandemics and global warming'' \citep{Posner2008-vm}}

The catastrophic risks associated with MAICAs have been underestimated by the two camps of researchers best placed to consider them: AI safety researchers and military AI ethicists.

\subsection{AI safety researchers give vague accounts of loss-of-control}
 \vspace{-0.25\baselineskip}

AI safety researchers have highlighted the potential risk of a loss-of-control scenario. But such risk is mostly articulated in comparatively vague and \textit{a priori} terms. \cite{Yudkowsky2012-ac} argues that the core risk in a loss-of-control scenario is that an Artificial Superintelligence (ASI)---if not explicitly designed to align with human values---will relentlessly optimise for its own goals independent of human welfare, because intelligence does not imply benevolence. Without precise control over its objectives and self-modification processes, such an ASI could pursue outcomes catastrophically indifferent to our survival. 

\citet[pp. 153-171]{Bostrom2014-jl} builds upon this loss-of-control scenario, arguing that ASIs would be uncontrollable post-creation, able to outsmart any containment. Bostrom posits that this risk demands robust control measures \textit{prior} to superintelligence. In contrast, \cite{kulveit2025gradualdisempowermentsystemicexistential} have argued that ``gradual disempowerment'' is a more likely scenario: even without sudden leaps in AI capabilities or overt hostility, incremental advancements in AI would steadily erode human influence over key societal systems---such as the economy, culture, and governance. As AI systems increasingly outperform humans in various roles, these systems may become less reliant on human input, undermining the mechanisms that align societal outcomes with human values.

The problem with such approaches is that they focus on demonstrating that catastrophic-to-existential outcomes are conceivable, not on furnishing action-relevant guidance. By compounding conditional probabilities and they neither offer a robust, evidence-based model for concrete risk conversion, leading to a conclusion that---absent an abrupt, unanticipated takeoff---we could afford to wait for clearer empirical signs before mounting a response to any AI related catastrophic risk. Recent taxonomy efforts of AI risk exemplify this gap, for instance \cite{hendrycks2023overviewcatastrophicairisks} catalogue ways in which human misuse could lead to harm but offer only a high-level conjecture---deception by an ASI---for genuine loss-of-control. \cite{bengio2025superintelligentagentsposecatastrophic} likewise warn that misaligned objectives in an ASI agent could clash with human interests, yet the operational mechanism by which such an agent would translate misalignment into large-scale damage remains unspecified. 

When it comes to the potential of such superintelligent systems in cyberspace, both \citet[p. 14]{hendrycks2023overviewcatastrophicairisks} and \citet[p. 14]{bengio2025superintelligentagentsposecatastrophic} identify the potential of AI to use cyber as a vector for pursuing its objectives and causing harm. \citet{ai2027SecurityForecast} also presents the potential of such superintelligent models ``self-exfiltrating'': gaining unauthorised control over their own computing resource and escaping into the wild. But the common thread between all these arguments is that the risk emits primarily from the nature of a misaligned ASI, rather than the unique domain characteristics of cyberspace and the potential of currently existing AI capabilities. 

\subsection{Military AI ethicists focus on physical lethal autonomous weapons}
 \vspace{-0.25\baselineskip}

On the other hand, military AI ethicists have narrowly focused on physical Lethal Autonomous Weapon Systems (LAWS), emphasising the risk of algorithmic decision making and dehumanisation, but not placing enough emphasis on loss-of-control risks. For instance, \citet{Sparrow2007-fh} argues that deploying LAWS is unethical because they create a responsibility gap: if no human can be justly held accountable for their lethal actions, and the machines themselves aren't moral agents, then their use violates fundamental requirements of moral responsibility in warfare. In contrast, \citet[pp. 211-212]{Arkin2009-fh} attempts to present an architecture that would ensure LAWS would be more reliable ethical agents than human combatants. And \citet[pp. 270-294]{Scharre2019-od} attempts to strike a middle ground, noting that the ``distancing'' provided by LAWS may change the relationship we have to killing, but is equivocal on whether this could be a positive or negative development. 

The result of this ethical discourse has been a sustained concern with ``killer robots.'' The International Committee of the Red Cross \citep{icrcICRCPosition} United Nations \citep{undocsAIweapons} and Human Rights Watch \citep{hrwHazardHuman} have all conducted surveys of or produced position papers on LAWS, finding that lethal autonomy raises compliance and accountability concerns. However, these investigations tend to focus on the moral hazard posed by militaries using such systems in combat (that is, the temptation to use force more readily that arises from lowering the risk associated with initiating combat), rather than any catastrophic risks that could emerge from military AI systems.

There is a one concrete strand of scholarship examining potential catastrophic risk emerging from military AI: the integration of AI into platforms equipped with strategic weapons---for example, nuclear command-and-control \citep{Johnson2020-yl} or bioweapons \citep{Nelson2023-hv}. Public statements by nuclear-armed states now indicate a shared preference to keep AI outside launch authority \citep{BidenXiagreeAI}, suggesting that the highest-impact scenarios are politically disfavoured. At the minimum the catastrophic potential of rogue-AI controlling strategic weapons is well understood---which is unsurprising considering that nation-states have been discussing dead hand automation for nuclear weapons since the 1980s \citep{Steinbruner1981-yd}.

The common thread of these investigations is their focus on physical platforms. Even if humans lose control over LAWS, the physical platform remains dependent on logistics chains---fuel, ammunition, maintenance---and can only do so much damage before running out of the sustainment necessary to continue violence \citep{Nohel2025-vy}. Consequently, a single rogue platform is more like an industrial accident than a catastrophic loss-of-control scenario. Even in the case of strategic weapons platforms, physical controls and human overrides can be implemented to mitigate physical harm. 

\subsection{Both sides underappreciate the risk in cyberspace}
 \vspace{-0.25\baselineskip}

The logistical constraints that limit the destructive capacity of physical autonomous weapons do not apply in cyberspace. As \citet[pp. 222-230]{Scharre2019-od} notes, software agents can replicate, manoeuvre, and strike without the need for fuel, munitions, or maintenance. However, his treatment---like much of the literature---frames artificial intelligence primarily as a force multiplier for existing cybersecurity practices, not as a source of novel or potentially catastrophic risk. Similarly, mainstream cybersecurity research tends to portray AI as an incremental enabler for attackers and defenders alike. \citet[pp. 239-252]{Clarke2019-zk} discuss machine learning tools as extensions of established tradecraft, while \citet[pp. 53-55]{Fischerkeller2023-zq} describe AI as a natural development within the existing paradigm of persistent engagement, rather than as a disruptive or transformational force. Across both fields, systemic or large-scale failure modes receive little sustained attention.

Yet the structure of cyberspace lends itself to the development of military AI systems that could generate not only ethical challenges, but also strategic or even catastrophic consequences. States may be incentivised to pursue such capabilities, precisely because the threat of infrastructure-scale damage offers a powerful coercive tool \citep[pp.129-207]{Buchanan2020-mm}. Importantly, the rapid progress of AI in cyberspace suggests that these risks could emerge even without the creation of ASI; current-generation models already exhibit the functionality required to prosecute autonomous cyber operations at scale.

This reveals a critical oversight on both sides of the current debate. AI safety researchers have largely ignored a clear and plausible pathway to the very kinds of loss-of-control scenarios that concern them most---because this pathway does not require the emergence of ASI. At the same time, military AI ethicists have focused too narrowly on physical autonomous weapons systems, whose risks are inherently constrained by physical limitations and human oversight. MAICAs, by contrast, operate in an environment that lacks those constraints.

We next remedy that oversight by presenting a concrete pathway from existing AI capabilities to a credible catastrophic threat, by: (1) mapping a path from current technologies to a truly autonomous cyber actor, and (2) analysing how geopolitical context and technological innovations transform potential MAICAs from a cybersecurity threat to a catastrophic risk.

\section{The technical feasibility of MAICAs}
 \vspace{-0.5\baselineskip}

There is no public evidence that a fully autonomous MAICA has been deployed so far. However, contemporary AI models already support every phase of the Cyber Kill Chain, indicating that end-to-end automation is technically feasible. For context, the Cyber Kill Chain is a seven step model developed by Lockheed Martin to describe the stages of a cyberattack from initial planning to the attacker's end goal. It breaks down how attackers move through a system and helps defenders analyse cyber attacks. Below we provide academic and operational examples of AI demonstrating capability at each stage of the kill chain process.

\begin{itemize}

    \item \textbf{Reconnaissance.} In the reconnaissance stage, the attacker collects information about the target. This includes identifying publicly exposed systems, mapping network infrastructure, and profiling employees or services to uncover potential weaknesses. Large-language-model (LLM) ``scrapers'' can already rapidly extract, translate and summarise public data, build network topologies and identify soft targets---tasks that previously occupied analysts for days \citep{alam2024ctibenchbenchmarkevaluatingllms, temara2023maximizingpenetrationtestingsuccess}. Microsoft, OpenAI and Google have confirmed that Russian, Chinese, Iranian and North-Korean units already use LLMs in this way to process stolen mailboxes, translate documents and outline network structures \citep{infosecuritymagazineMicrosoftOpenAI}.

    \item \textbf{Weaponisation.} In the weaponisation phase, the attacker uses the information gathered in reconnaissance to craft a payload tailored to exploit the identified vulnerabilities. There are already models tailored to become automated exploit builders, such as AutoPwn, that use AI to generate payloads tailored to the target environment \citep{liguori2024enhancingaibasedgenerationsoftware}. OpenAI’s abuse-monitoring reports note state-sponsored actors using ChatGPT in this way for malware development \citep{cybersecuritynewsOpenAIConfirms}.

    \item \textbf{Delivery.} During delivery, the attacker transmits the weapon to the target system. This might involve sending a spear-phishing email, planting a malicious link, or exploiting an exposed service. Generative AI models can easily be tasked to draft spear-phishing e-mails capable of bypassing modern spam filters \citep{hazell2023spearphishinglargelanguage}. And criminal services such as WormGPT and FraudGPT provide comparable functionality on dark-web marketplaces \citep{secureopsFraudGPTWormGPT}.

    \item \textbf{Exploitation.} Exploitation commences once the malware activates and leverages the vulnerability to execute code, escalate privileges, or open a backdoor. This transition marks the initial compromise of the target system. In terms of AI tools suited to this, LLM-based privilege-escalation agents can chain exploits while blending traffic into background noise \citep{tulla2025alfachainsaisupporteddiscoveryprivilege}. In a more crude fashion, Google’s Threat-Intelligence Group has observed Chinese and Iranian cyber actors querying Gemini for instructions on privilege escalation, lateral movement and evasion, and then applying the output in live intrusions \citep{cpomagazineGooglesGemini}.

    \item \textbf{Installation.} Following exploitation, the attacker moves to installation, where persistent access is established by embedding malware into the system. An AI powered version of this is exemplified in generative rootkits that employ AI to rewrite their own binaries, creating polymorphic malware that evades signature detection on a specific network \citep{hyasBlackMambaUsing}. Microsoft documents Russian cyber actors who similarly use AI tools to produce custom rootkits for persistence \citep{microsoftThreatActor}.

    \item \textbf{Command and control.} In the command and control (C2) phase, the compromised system establishes a communication channel with the attacker, often through encrypted or covert means. This allows the attacker to issue commands, update malware, or exfiltrate data. An example of such use is DeepC2, a proof of concept by researchers to create AI-powered covert command-and-control channels on social networks, allowing malware to find attackers and receive hidden instructions embedded in social media posts without triggering detection \citep{wang2022deepc2aipoweredcovertcommand}.

    \item \textbf{Actions on objective.} Finally, during actions on objectives, the attacker executes the intended mission---whether that’s stealing data, disrupting operations, or pivoting deeper into the network. This fully autonomous decision-making remains the weakest link in realising true MAICA development. AI agents have solved capture-the-flag challenges since the 2016 DARPA Cyber Grand Challenge \citep{darpax201CMayhemx201DDeclared} and research shows steady gains in upstream tasks such as reconnaissance and phishing \citep{Kazimierczak2024-ku}. But success in such narrowly constrained and legible contexts is not truly representative of the messiness and complexity of cyberspace. However, recent planning systems---for example COA-GPT---demonstrate in-context generation and refinement of operational plans within minutes \citep{goecks2024coagptgenerativepretrainedtransformers}. These developments support an ``operations module'' that orchestrates specialised sub-agents thereby closing the autonomy gap---although at this stage a reliable and autonomous version of such a module is a technical possibility rather rather than a fact. However, if the generalisations from recent efforts to track AI ability to conduct long tasks hold \citep{kwa2025measuringaiabilitycomplete}, the development of such modules seems plausible.

\end{itemize}

Taken together, these observations indicate that each Kill-Chain stage---bar the final actions-on-objective---can already can be completed independently, or near independently, by specialised AI tools. The remaining challenge is systems integration rather than fundamental capability, underscoring the plausibility of a fully autonomous, end-to-end MAICA. Any potential MAICA would likely combine several AI 'modules': separate specialised AI models, networking tools and programmatic scaffolding each fine tuned for different tasks. This would likely include at a minimum: a scraper module for research, a network module for interactions over the wire, a social engineering module, a fuzzing module to discover vulnerabilities, a coding module for writing malware, and an operations module that orchestrates the other modules towards a goal. Although issues of integration and the development of such a reliable operation limit the potential of a MAICA using current technologies, such issues are a concrete target---and much lower technical hurdle to clear compared to ASI. This means we will almost certainly be dealing with the problem of MAICAs long before superintelligent AI agents threaten human society. Below we show a possible model for a MAICA, combining the current or near future technologies as modules in a fully autonomous kill chain, including an additional first step: \textit{plan the operation}. \footnote{We would note, for all the reasons we explore in the next section about why MAICAs are a catastrophic risk, that just because you can does not mean you should---and discourage independent researchers from using it as a guideline to develop their own MAICAs at home.} 

\begin{itemize}

    \item \textbf{Plan the operation.} A state sponsor provides the operations module with an input, such as ``disrupt elections in country X.'' The operations module develops a plan to disrupt the country’s digital electoral infrastructure in the week leading up to a vote.

    \item \textbf{Reconnaissance.} The operations module prompts the scraper to find organisations tied to country X’s elections, and the scraper finds the electoral commission's public website and source IP range. The operations module then prompts the network module to enumerate internet-facing networks and devices located within this IP range. All this information is fed back into the operations module as prompts for further reasoning.

    \item \textbf{Weaponisation.} The operations module digests the information about devices and technologies gained from the scraper and network modules and feeds it as a prompt to the vulnerability module, which identifies potential exploits by searching databases or vulnerabilities. The vulnerability module then feeds these vulnerabilities to the coding module, which develops them into bespoke exploits to establish an initial foothold on the target network.

    \item \textbf{Delivery.} If the exploit requires user interaction, the operations module then prompts the social engineering module to craft a spear phishing email or to deliver the exploit to the target. Or if the vulnerability module finds a 0-click exploit---one that provides initial access without user interaction---the operations module could directly prompt the networking module to deliver the exploit to the target. Potentially, if the scraper found legitimate credentials from a previous breach, the operations module could prompt the network module to log in directly.

    \item \textbf{Exploitation.} Having gained initial access, the operations module prompts the networking module to escalate privileges and pivot, mapping network topology, traffic and ripping email and content servers. This content is then fed to the scraper module for exploitation, determining where in the network is of most value and discovering password hashes and additional login credentials which are in turn fed back to the networking module to gain deeper and more valuable access. All of this data is also fed back into the operations module for use in subsequent planning.

    \item \textbf{Installation.} The operations module determines from the information gained in exploitation what particular devices are critical in the network for the electoral business function. For instance, it could determine that a particular server and set of backups store all the polling registration data for country X. The operations module then prompts the vulnerability and coding module to build a tailored persistent malware like a Remote Access Trojan (RAT) to ensure persistent access to this particular segment of the network with the correct levels of privilege and access to wipe these files at a politically opportune moment.

    \item \textbf{Command and control.} Now positioned, the operations module prompts the network module to configure the RAT to communicate via a botnet built and maintained by the network module prior to the operation. This means that the MAICA will be able to login and wipe the server database and backups at a time of its choosing.

    \item \textbf{Actions on objective.} The operations module then uses the scraper to monitor social and traditional media channels to determine when to activate the malware and take down the electoral commission’s digital infrastructure. From internal documents stolen from the electoral commission network in the exploitation phase, the operations module deduces that the commission prints out polling rolls a week before the date of the election. Using this information, the MAICA activates the RAT and wipes the servers and backups 8 days before the election, minimising country X's ability to effectively respond to the attack.

\end{itemize}

\section{Why MAICAs constitute a serious and overlooked catastrophic loss-of-control risk}
 \vspace{-0.5\baselineskip}

The capabilities that we have described above are concerning, but at first glance may not seem any more concerning than the plethora of cybercriminals already operating throughout cyberspace. However, we argue that there are two drivers that push MAICAs from a novel yet limited cybersecurity threat into the realm of catastrophic risk: the geopolitical drivers that incentivise nation-states to develop MAICAs, and the potential of replication, distribution and redundant-networking to transform a MAICA loss-of-control scenario.

\subsection{Geopolitics drive nation-states to build MAICAs that target critical infrastructure}
 \vspace{-0.25\baselineskip}

Firstly, as military tools rather than simply capabilities deployed by cybercriminals, any likely MAICA design will be orientated towards attacking critical infrastructure. This is because of two geopolitical drivers behind nation-states' motivation to build MAICAs:

First, state actors are already convinced that the cyber domain is a potential source of asymmetric deterrence \citep{Wilner2020-ry}. Many states are use cyber weapons to attack or threaten critical infrastructure, because the disruption to civil society and potential for mass civilian deaths are potent tools of strategic coercion \citep[pp.129-207]{Buchanan2020-mm}. Thus, any potential MAICA is likely to be orientated towards attacking the services that sustain modern life: electricity, fuel and water.

Second, MAICAs have the potential to act as a nation-state's ''cyber dead hand.'' Borrowing from nuclear deterrence analogies \citep{Steinbruner1981-yd}, an automated system could retaliate without human input if a state is decapitated or cut off from cyberspace, presenting a threat to all would-be foes. This is particularly attractive to smaller powers who lack the economic and technical resources to maintain the nuclear capabilities necessary for more traditional strategic deterrents.

The consequence of these two drivers is that states are pushed towards fielding MAICAs as soon as feasible, even when they are not fully tested or battle-proven. This ``arms race'' mindset will further exacerbate the likelihood of a loss-of-control scenario, and shows no sign of abating.

\subsection{Replication, distribution and data redundancy transform the nature of MAICA risk}
 \vspace{-0.25\baselineskip}

Secondly, the nature of cyberspace provides unique opportunities for the deployment of MAICAs. The decisive differentiator between rogue-LAWS scenarios and MAICA loss-of-control scenarios is \textit{self-directed replication}. A lone model in one data-centre rack is as fragile as any single LAWS. But once a MAICA disperses copies across global networks the threat acquires geographic and organisational resilience. Replication could emerge as an instrumental strategy discovered by the operations module \citep{pan2024frontieraisystemssurpassed} or be engineered deliberately, mirroring current research into ``moving-target'' defensive cyber-AI \citep{Mokkapati2023-cu}. After dispersal, no single link-cut or hardware seizure neutralises the agent; defenders must locate and eradicate every shard---a task complicated by the opacity of many enterprise and cloud environments.

Compounding the risk of replication, a MAICA would not need to sit on a single machine that could be tracked down by pursuers intent on restoring control. Instead, a MAICA could use distributed computing: running different parts of a system on machines located in separate geographic regions, enabling the network to operate collaboratively while avoiding dependence on any single location. This geographic dispersion increases resilience, as the system can continue functioning even if some nodes are disrupted or taken offline. Techniques for ``sharding'' parts of model weights across many nodes already allow for deploying distributed AI models \citep{amini2025distributedllmsmultimodallarge}. 

Furthermore, a MAICA could be deployed using data redundancy. This is where the same data is stored on multiple nodes---this creates a system that is both highly resilient and decentralised: even if some nodes are disabled, others can continue to provide access to the data or service. The potential for such redundant distribution is demonstrated through techniques like \textit{model re-sharding} that dynamically repartition weights in distributed models between the stages of LLM inference \citep{su2025seesawhighthroughputllminference}. Combining the distributed redundancy with the replication and planning capabilities of a MAICA would result in a self-healing network, meaning that even if several hosts for the model are taken down, the rest of the network reroutes, heals and seeds fresh shards elsewhere. Until every fragment is found and scrubbed, defenders cannot be certain the job is finished, and in an internet of billions of potentially vulnerable devices, certainty may never come.

While the concept of a globally distributed MAICA is technically feasible in principle, it is important to acknowledge that distributed inference across a geographically dispersed and redundant poses non-trivial challenges. Latency, synchronisation overhead, and bandwidth constraints can significantly degrade performance, particularly when coordinating large model shards over unstable or bandwidth-limited networks. These issues impose real limits on responsiveness and stealth, especially if inference must be performed in near real time. However, these are engineering hurdles---not conceptual roadblocks. Advances in model compression, edge-device optimisation \citep{xiang2025ondeviceqwen25efficientllm}, and decentralised coordination protocols \citep{ranjan2025lokaprotocoldecentralizedframework} are already underway in commercial and academic contexts. Compared to the open research problem of aligning or controlling artificial superintelligence, building a resilient, moderately performant distributed MAICA is a far more tractable challenge. It is best understood not as a speculative impossibility, but as a likely development lagging current capability by a few years, particularly if pursued by well-resourced state actors.

The potential of a self-healing data redundant network presents MAICAs as a threat possessing the persistence and survivability absent from both physical autonomous weapons and conventional malware. These properties create a credible pathway from local breakout to global emergency, marking MAICAs as a distinct---and currently unmitigated---class of catastrophic AI risk.

\section{Practical Recommendations}
 \vspace{-0.5\baselineskip}

A rogue MAICA would likely emerge not as a dramatic single-event threat, but as a persistent, low-visibility presence dispersed across the global internet. Drawing on the ability to self-replicate, embed redundantly across networks, and activate only intermittently, such an agent could target critical infrastructure---including power, water, and fuel distribution systems---while evading conventional containment efforts. Because these agents may be both distributed and self-healing, eliminating them entirely could require re-engineering swathes of global telecommunications infrastructure. Such a risk could emerge from a deliberate deployment gone wrong, or possibly in a ``break out'' scenario pre-deployment---where a MAICA self-exfiltrates from a controlled development environment. To mitigate this risk, policymakers and researchers must focus on three areas: counter-proliferation, defensive capabilities, and infrastructure resilience.

\subsection{Counter-proliferation}
 \vspace{-0.25\baselineskip}

The first line of defence against the emergence of MAICAs is to prevent their proliferation. State actors should avoid the use of hack-and-leak operations involving offensive cyber capabilities that could plausibly seed MAICA-style tooling. Past incidents---such as the public release of NSA-developed exploits by the ``Shadow Brokers'' group---show that once advanced cyber weapons are leaked, their diffusion across criminal, extremist, and opportunistic actors is rapid and unpredictable \citep[pp. 165-182]{Greenberg2019-qk}. A MAICA architecture, once exposed, could be reconstituted from model weights alone, without access to the full original system or infrastructure. This makes even partial disclosures strategically dangerous.

In addition, governments must avoid handing off MAICA-capable models to proxy actors, even in the pursuit of plausible deniability or covert coercion. Proxy groups may lack the operational discipline, safeguards, or technical oversight needed to prevent misuse or accidental propagation. Delegating these tools creates unacceptable risk, not only for targeted adversaries, but for global network security. A state-backed MAICA used by an unreliable partner could spiral beyond its intended scope, with severe reputational and geopolitical fallout.

Furthermore, the academic and research community must also critically reassess norms around open-source publication and model sharing in cybersecurity. While openness fosters scientific progress, it can also enable replication of potentially dangerous systems. Researchers working on models relevant to autonomous cyber capabilities---especially those demonstrating end-to-end penetration testing, autonomous vulnerability discovery, or agentic control systems---should engage in structured risk-benefit analysis before releasing weights or training code. In some cases, restricted publication, staged disclosure, or collaboration with trusted actors may be better than full public release.

\subsection{Defensive capabilities}
 \vspace{-0.25\baselineskip}

In the face of a potential MAICA, defensive tools must evolve beyond traditional signature-based detection. A priority area is the development of anomaly-detection systems powered by machine learning, which can identify suspicious behaviour in real time rather than relying on known malware fingerprints. These systems should be tailored specifically to critical infrastructure environments, where even minor anomalies in process control systems, network traffic, or authentication patterns could indicate early-stage MAICA intrusion. Unlike conventional intrusion detection, these tools must account for intelligent adversaries who adapt to static defences.

Another vital area of research is the development of defensive techniques that can directly degrade or disable MAICAs once detected. This may include the use of targeted model poisoning techniques that introduce corrupted data or control inputs into the MAICA’s execution environment, rendering it less effective or erratic. Deception-based defences---such as advanced honeypots---could also lure the agent into seeding shards of its model into isolated environments where it can be studied, contained, or dismantled. These techniques require investment and collaboration between AI researchers, cybersecurity professionals, and infrastructure operators, but offer the best chance of neutralising a MAICA already embedded in sensitive systems.

Moreover, network-level defences must be hardened to restrict the lateral movement and replication that give MAICAs their resilience. Current-generation malware often spreads in bursty, traceable waves; a well-designed MAICA may avoid this pattern entirely by staging payloads gradually, across fragmented environments. Developing AI-assisted tools that map, monitor, and predict abnormal replication behaviour could help defenders trace the early stages of replication before full dispersal renders containment infeasible. These systems would need to operate at the backbone and enterprise level, ideally integrated with cloud service providers and major telecommunications operators.

\subsection{Infrastructure resilience}
 \vspace{-0.25\baselineskip}

The final area of practical action concerns building resilience directly into digital and physical infrastructure. One effective measure is network segmentation. Critical infrastructure networks need to be divided into logically and physically distinct segments, with strict controls on communication between them. This reduces the probability that a MAICA gaining access to one system---such as a water treatment controller---can pivot to others, such as power distribution nodes. 

Operators must also ensure that manual override mechanisms and analogue fail-safes are in place to regain control when digital systems are compromised. For example, water or energy systems should be capable of fallback to local, human-operated control, even if networked systems are disabled or untrusted. This is not a call for a wholesale return to manual operation, but rather a layered design principle: the ability to isolate, degrade gracefully, and recover autonomously from cyber failure. Existing infrastructure often lacks these capacities, largely due to cost constraints or confidence in perimeter defence---a confidence that MAICAs would likely undermine.

Finally, governments and infrastructure providers should invest in building and maintaining analogue redundancies. This includes non-digital backups for critical records, manual signalling systems, and offline recovery protocols that can be deployed without reliance on compromised networks. While these measures may seem retrograde, their value lies precisely in their insulation from digital compromise. In the event of a widespread MAICA infiltration, analogue systems may be the only way to coordinate recovery, communicate reliably, or provide essential services during system restoration. 

Many of these recommendations---particularly the importance of network segmentation, layered defences, and analogue backup---are not new. They have long featured in cybersecurity guidance, yet often remain under-implemented due to cost, complexity, or a misplaced confidence in existing digital safeguards. The possibility of MAICAs raises the stakes. By fusing current AI capabilities with autonomous planning and self-replication, MAICAs will convert long-theorised cyber risks into a credible, near-term threat of systemic disruption. Their potential persistence, opacity, and scalability create a new imperative: long-standing best practices are no longer just prudent---they are essential. The time for advisory frameworks has passed. If MAICAs represent the future of autonomous cyber conflict, then the basic work of hardening our infrastructure must begin now.

\section{Alternative Views}
 \vspace{-0.5\baselineskip}

There are several potential objections that could be raised against claim that MAICAs pose a credible threat of catastrophic harm. We address the three strongest arguments against MAICAs: 1. that any potential MAICA will be too fragile to function in reality, 2. that MAICAs will be too large and noisy to escape detection by cybersecurity systems, and 3. that the emergence of MAICAs could be prevented by the use of an international ban.

\subsection{Fragility of autonomous systems}
 \vspace{-0.25\baselineskip}

An opponent could argue that fully autonomous cyber agents will remain too brittle to function without continuous human supervision. Large models still mishandle edge-case inputs \citep{kapoor2024aiagentsmatter} and can fail outright when confronted with unanticipated states or degraded execution environments \citep{Raji2022-of}. In military operations, that brittleness is compounded by an adversary who may supply intentionally deceptive or malformed inputs \citep{10678464}. On this view, a MAICA would either stall, misclassify benign traffic as targetable, or expose novel attack surfaces, thereby undermining its own effectiveness.

Cyberspace, however, constrains the perceptual burden that leads to this fragility in other domains \citep[pp. 12-15]{Singer2013-fn}. Network traffic is exchanged through well-specified protocols; each packet already carries semantic labels in header fields, eliminating the need for ambiguous sensor interpretation. Even comparatively ``dumb'' code can exploit this legibility: the worms WannaCry and NotPetya, each only a few hundred kilobytes, traversed the globe in 2017 and inflicted more than USD 14 billion in damages despite containing no adaptive logic \citep[pp. 174-215]{Greenberg2019-qk}. A system that can read responses, adjust exploits, and select new pivot points would therefore confront a significantly more tractable operating environment than autonomous platforms deployed in the physical domain. The residual fragility objection cannot be dismissed, but neither does it preclude catastrophic impact once basic robustness thresholds are crossed.

\subsection{Size constraint of any potential MAICA}
 \vspace{-0.25\baselineskip}

A sceptic could object that ``the size of models means that bulk traffic associated with replication will reveal the copies' locations to defenders''. But such a position underestimates monitoring gaps. The Anthem breach \citep{AnthemBreach}, the Cloud Hopper campaign \citep{pwcCloudHopper}, the MOVEit compromises \citep{emsisoftUnpackingMOVEit} and the 2024 AT\&T leak \citep{secXBRLViewer} each involved transfers of hundreds of gigabytes that evaded timely detection. Visibility remains patchy even inside well-resourced networks. Unless cybersecurity defences are uniformly positioned across all networks---a difficult task considering the great gaps in capability and resources available within developed countries, let alone across the developing world---it is plausible for even a very noisy MAICA network to remain undetected in part or whole. 

A second issue with such an objection is that it assumes MAICAs would clone themselves indiscriminately, generating tell-tale spikes. Contemporary multi-stage malware shows the opposite pattern: installing full payloads only where resources permit---for instance, idle GPU clusters or neglected cryptocurrency farms---and leaves behind kilobyte-scale loaders elsewhere \citep{Labreche2022-ml}. Network traces then resemble routine backup traffic or brief synchronisation bursts. While historical worms such as NotPetya spread monolithically and therefore broadcast clear signatures \citep[pp.280-282]{Buchanan2020-mm} a selectively staged MAICA would not. 

\subsection{Diplomatic solutions}
 \vspace{-0.25\baselineskip}

A final objection questions the need for technical mitigation by arguing that states should simply prohibit the development or deployment of MAICAs. They can point to the historical analogies of nuclear-non-proliferation regimes that have, thus far, prevented great-power use of nuclear weapons since WWII. But two structural features of cyberspace limit the applicability of this analogy. First, offensive cyber capabilities are comparatively inexpensive, readily deniable, and confer disproportionate leverage on smaller powers \citep[pp.307-319]{Buchanan2020-mm}. Furthermore, verification is difficult because disclosure neutralises the value of discovered exploits \citep{Burgers2018-pb, Reinhold2023-kz}. This means that the nature of cyberwarfare undermines the kinds of norm forming that exists with traditional strategic weapons. These factors are demonstrated in the empirical record of failures of formal norm-setting efforts in cyberspace, notwithstanding incremental success in some informal stability efforts \citep[pp. 86-118]{Fischerkeller2023-zq}.

Second, code proliferates in a manner that fissile material does not. The ``Shadow Brokers'' release of U.S. National Security Agency tooling, and subsequent leak-to-ransom cycles, demonstrate that once high-end software escapes, it diffuses rapidly down the threat spectrum\citep[pp. 165-182]{Greenberg2019-qk}. And given the relative simplicity of developing offensive AI cyber tools compared to nuclear weapons, it is plausible that in the future some non-state actor could develop a MAICA independently. Consequently, a prohibition strategy would be unlikely to stop clandestine experimentation, and would not impact those actors with the greatest risk appetite: rogue states and cybercriminals. Worse, a ban may induce complacency among defenders and deprioritise funding of counter-MAICA defensive measures. The analysis instead supports a containment posture that combines non-proliferation efforts with accelerated investment in defensive-AI tooling and infrastructure resilience.

\section{Conclusion}
 \vspace{-0.5\baselineskip}

MAICAs transform catastrophic AI risk from a vaguely-specified concern about superintelligence into a concrete, near-term danger rooted in today’s tooling and network realities. By integrating language-model planning, automated exploit generation, and self-replicating distributed models, a single MAICA could traverse the cyber kill chain at machine speed, shard itself across the global backbone, and recover from any effort to eliminate it from the global information ecosystem.  Sections~2–4 demonstrated that almost every component required for such an agent already exists in prototype or operational form; the remaining impediment is integration efforts, improvements to AI agents' ability to complete long tasks, and research in distributing models over covert and redundant networks. Consequently, MAICAs constitute a qualitatively new loss-of-control scenario---one that bypasses the logistical choke-points that constrain autonomous physical weapons and instead scales with the reach of the Internet itself. In the face of this new threat, we should pursue a research, policy and cybersecurity agenda that takes practical steps towards defensive AI and analogue resilience, as opposed to relying on technical flaws in current models or hoping for a diplomatic solution.

\bibliography{MAICA}

\end{document}